\documentclass[a4paper]{panle}
\usepackage{cite}
\usepackage{wrapfig}
\usepackage{graphicx}
\usepackage{amssymb}
\usepackage{amsfonts}
\usepackage{amsmath}
\usepackage{longtable}
\usepackage{rotating}
\usepackage{lscape}
\usepackage{epsfig}
\usepackage{multirow}
\usepackage{color}

\usepackage{lineno}

\originalTeX
\begin{document}

\title{Optimisation of the Accelerator Control by Reinforcement Learning: A Simulation-Based Approach}

\maketitle
\authors{A.\,Ibrahim\,$^{a,}$\footnote{E-mail: aibrahim@hse.ru},
D.\,Derkach\,$^{a,}$\footnote{E-mail: dderkach@hse.ru}, 
A.\,Petrenko\,$^{b,}$\footnote{E-mail: A.V.Petrenko@inp.nsk.su}, 
F.\,Ratnikov\,$^{a,}$\footnote{E-mail: fratnikov@hse.ru}
M.\,Kaledin,$^{a,}$\footnote{E-mail: mkaledin@hse.ru} }
\from{$^{a}$\,Laboratory of Methods for Big Data Analysis, HSE University, Moscow, Russia}
\from{$^{b}$\,Budker Institute of Nuclear Physics, Novosibirsk, Russia}


\begin{abstract}
Optimizing accelerator control is a critical challenge in experimental particle physics, requiring significant manual effort and resource expenditure. Traditional tuning methods are often time-consuming and reliant on expert input, highlighting the need for more efficient approaches. This study aims to create a simulation-based framework integrated with Reinforcement Learning (RL) to address these challenges. Using \texttt{Elegant} as the simulation backend, we developed a Python wrapper that simplifies the interaction between RL algorithms and accelerator simulations, enabling seamless input management, simulation execution, and output analysis.

The proposed RL framework acts as a co-pilot for physicists, offering intelligent suggestions to enhance beamline performance, reduce tuning time, and improve operational efficiency. As a proof of concept, we demonstrate the application of our RL approach to an accelerator control problem and highlight the improvements in efficiency and performance achieved through our methodology. We discuss how the integration of simulation tools with a Python-based RL framework provides a powerful resource for the accelerator physics community, showcasing the potential of machine learning in optimizing complex physical systems.

\end{abstract}

\vspace*{6pt}
\noindent

\section*{\uppercase{Introduction}}
\label{sec:intro}

Physics problems are inherently complex and demand significant effort to identify optimal parameters that maximize performance and achieve desired outcomes. Among the tools that have advanced our ability to tackle these challenges are particle accelerators, complex devices that employ electromagnetic fields to accelerate charged particles to high speeds and energies. These accelerators play a pivotal role in a wide range of scientific and industrial applications, including particle physics research, materials science, and medical technologies~\cite{Lee2020}.

The functioning of modern particle accelerators requires a sophisticated setup and fine-tuning of many individual beamline elements like dipole and quadrupole magnets, accelerating cavities, etc. As an example of such beamline in this paper we use the positron transport beamline between the linear accelerator and the damping ring of the VEPP-5 Injection Complex~\cite{Blinov:2014hka, rybitskaya2008production} at the Budker Institute of Nuclear Physics (BINP). The VEPP-5 injection complex is the source of all electron and positron beams used in the VEPP-4 and VEPP-2000 collider experiments at BINP. The beam transmission efficiency of this particular beamline is important as one of several bottlenecks limiting the luminosity of collider experiments.

Optimizing beamline performance is essential for achieving the desired experimental outcomes and minimizing resource expenditure. While traditional optimization methods often rely on manual tuning and heuristic approaches, recent advancements in machine learning have introduced more systematic and efficient alternatives. Reinforcement Learning (RL), in particular, has demonstrated significant potential in solving complex optimization problems across various domains. For instance, RL has been successfully applied to master complex tasks such as playing Atari games \cite{Mnih2015} and optimizing strategies for the game of Go \cite{Silver2017}. Additionally, RL has shown promise in robotics for optimizing control policies and motion planning \cite{Kober2013}, as well as in combinatorial optimization problems like the traveling salesman problem and vehicle routing \cite{Li2020, Bello2016}. Recent advancements, such as DeepSeek, have further demonstrated RL's capability to tackle large-scale optimization problems in diverse fields, including industrial automation and resource allocation \cite{DeepSeek2023}. Studies have shown RL capability to enhance beamline performance in accelerators by improving control stability, beam quality, and operational efficiency while reducing costs \cite{cern_rl2020, bayesian_fel2020, beamline_rl2021}.

In this work, we aim to leverage RL to address the challenges of beamline optimization. Our goal is to develop a system that acts as a co-pilot for physicists, providing intelligent suggestions to enhance beamline performance and reduce the time and effort required for manual tuning. By integrating RL with simulation tools such as \texttt{Elegant}, our framework supports both accelerator physicists and machine learning researchers, offering a versatile setup for optimizing various beamline architectures. This approach bridges the gap between accelerator physics and machine learning, empowering both fields to achieve significant advancements.

In this work, we aim to leverage RL to address the challenges of beamline optimization. Our goal is to develop a system that acts as a co-pilot for physicists, providing intelligent suggestions to enhance beamline performance and reduce the time and effort required for manual tuning. To achieve this, we created a reinforcement learning environment integrated with the \texttt{Elegant} simulation tool. This environment allows researchers to explore and train RL agents on various beamline configurations and scenarios, making it a versatile resource for both accelerator physicists and machine learning researchers.

\section*{\uppercase{Literature Review}}
\label{sec:review}

The application of machine learning, particularly reinforcement learning (RL), in accelerator optimization has received considerable attention. Pang et al.~\cite{Pang2020} demonstrated the potential of deep RL combined with high-fidelity simulations for optimizing control policies in a particle accelerator. Their work focused on a section of the accelerator, showcasing better-than-human performance in terms of beam current and distribution. Despite the promise, the study highlighted challenges such as the need for extensive data and the high computational cost of training.

Kaiser et al.~\cite{Kaiser2024} introduced Cheetah, a high-speed, differentiable simulation tool designed to bridge the gap between machine learning and accelerator physics. By significantly reducing simulation times, Cheetah enables efficient RL training and gradient-based optimization for beamline tuning. The tool supports a wide range of applications, from Bayesian optimization to neural network surrogate modeling, making it a versatile resource for the field.

Hirlaender et al.~\cite{Hirlaender2024} applied meta-reinforcement learning (Meta-RL) and model-based RL (MBRL) to particle accelerator control. Using simulation environments such as MAD-X, they addressed the challenges of partial observability and model inaccuracies. Meta-RL facilitated rapid adaptation of trained agents to real-world conditions, while MBRL demonstrated extreme sample efficiency through Gaussian Process-based Model Predictive Control (GP-MPC).

Kaiser et al.~\cite{Kaiser2024b} explored the use of Large Language Models (LLMs) for autonomous accelerator tuning. Their work demonstrated that LLMs could solve tuning tasks based on natural language prompts, providing an intuitive interface for specifying tuning goals. However, their performance lagged behind state-of-the-art algorithms like Reinforcement Learning Optimization (RLO) and Bayesian Optimization (BO). Positive aspects include the potential for straightforward deployment and enhanced human-machine collaboration. Negative aspects include the high computational cost, environmental impact, and limited current competitiveness with specialized optimization algorithms.

These advancements highlight the potential of RL, simulation-based approaches, and LLMs for optimizing particle accelerator systems. However, limitations such as simulation fidelity, data requirements, and real-world adaptation persist. The integration of tools like \texttt{Elegant}, Cheetah, and LLMs into RL frameworks represents a significant step forward, yet achieving seamless transfer between simulation and reality remains an active area of research.

\section{\uppercase{Methodology}}
\label{sec:Methodology}

To address the beamline optimization problem, we design a Reinforcement Learning (RL) framework where an \textbf{agent} interacts with an \textbf{environment} to iteratively learn optimal control policies. The interaction process is structured into \textbf{episodes}, with each episode consisting of several steps. At each step, the agent performs an action (adjusting the magnet current), receives feedback from the environment in the form of a state and reward, and progresses sequentially along the beamline until the end of the episode.


\subsection{Episodes \\ \\}

An episode begins with the agent positioned at the source of the beamline. The terms \textbf{action}, \textbf{state}, and \textbf{reward} will be explained in detail in Section 1.2. Environment. The agent:
\begin{itemize}\addtolength{\itemsep}{-2mm}
    \item Takes an \textbf{action} at each step by determining the current for the magnet at that position.
    \item Observes the \textbf{state}, which provides a description of the beamline after the action is applied.
    \item Receives a \textbf{reward} based on the effectiveness of its action in maintaining particle transmission and minimizing losses.
\end{itemize}

The agent continues this process for each magnet along the beamline. The episode concludes when:
\begin{itemize}\addtolength{\itemsep}{-2mm}
    \item The beam successfully reaches the target, completing the beamline.
    \item The particle count drops to zero, indicating a failure.
\end{itemize}

\medskip

\subsection{Environment \\ \\}

The environment represents the beamline simulation, built using our custom implementation integrated with the \texttt{Elegant} tool~\cite{Borland:2000gvh}. It provides a virtual testbed where the agent can explore and optimize beamline configurations safely. For each step in an episode, the environment simulates the beamline's response to the agent's action and provides feedback.

\medskip

\subsubsection{State \\[1ex]}

The state represents the beamline's condition at a specific step. It includes:
\begin{itemize}\addtolength{\itemsep}{-2mm}
    \item Beam dynamics parameters such as beta functions (\( \beta_x, \beta_y \)) and dispersion (\( \eta_x, \eta_y \)).
    \item Beam size (\( \sigma_x, \sigma_y \)) and aperture limits.
    \item Coupling factors and the particle count (\( N_p(i) \)) at a specific longitude \((i\)) in the beamline.
\end{itemize}
This state satisfies the Markov property~\cite{Sutton1998}, as confirmed by \texttt{Elegant} simulation outputs (manual Sections 4.1, 5.3, 5.6, 5.7, 8.2), which show that these parameters fully determine the beam’s evolution at each step~\cite{Borland:2000gvh}.

\medskip

\subsubsection{Reward \\[1ex]}

The reward function \(R(i)\) incentivizes efficient beam transport while penalizing particle losses. It is defined as:
\begin{equation}
R(i) = \Big[\big( N_{p}(i) - N_{p}(i = 0) \big) + \big( N_{p}(f) - N_{p}(i = 0) \big) - \sqrt{f^2 - j^2} \,\Big],
\end{equation}
where \( N_p(i) \) is the particle count at longitude \( i \), \( N_p(i=0) \) is the initial particle count, \( N_p(f) \) is the particle count at the target, \( j \) represents the longitude where all particles are lost, and \( f \) is the longitude at the target. 

The reward function consists of three terms that guide the agent to maximize particle transmission efficiency by minimizing particle losses. The first term, \( N_p(i) - N_p(i=0) \), represents the number of lost particles up until longitude \( i \), corresponding to a specific step in the episode and a specific magnet. This term penalizes the agent for any particles lost at intermediate steps. The second term, \( N_p(f) - N_p(i=0) \), measures the total number of lost particles over the entire beamline, encouraging the agent to keep as many particles as possible until they reach the target. Finally, the third term, \( -\sqrt{f^2 - j^2} \), penalizes the agent for incomplete episodes, where \( j \) is the first longitude at which all particles are lost and represents the distance between this point and the beamline end (target position), not a particle count, Its scale depends on positional loss patterns, not particle counts, across configurations. The goal here is to minimize this term by ensuring that all particles are successfully transported to the target, i.e., \( j = f \).

Thus, by maximizing the reward, the agent aims to minimize the three components: (1) the lost particles at each step, (2) the total number of lost particles along the beamline, and (3) the distance at which particles are lost before reaching the target. In other words, maximizing the reward corresponds to minimizing particle losses and ensuring all particles reach the target, optimizing the beamline's performance; we chose this multi-term design after testing simpler rewards (e.g., only current particles), which failed due to the non-greedy nature of beamline effects.

\medskip

\subsubsection{Action \\[1ex]}
The agent’s action space consists of controlling the currents of the magnets:
\begin{itemize}
    \item Quadrupole magnets: Continuous currents range between \([-20, 20]\).
    \item Dipole magnets: Continuous currents range between \([-0.05, 0.05]\).
\end{itemize}

\begin{figure}[h]
\begin{center}
\includegraphics[width=0.6\textwidth]{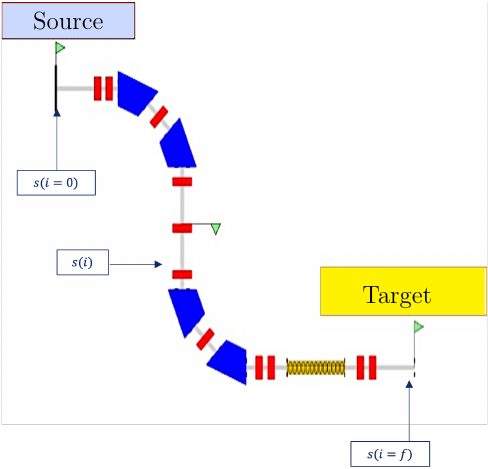}
\vspace{-2mm}
\caption{Schematic of the beamline, showing quadrupole magnets (red) and dipole magnets (blue). The beam starts at the source and terminates at the target. Sequential decision-making is employed to optimize magnet settings at each step.}
\end{center}
\labelf{fig:beamline}
\vspace{-5mm}
\end{figure}

\medskip

\subsection{Agent \\ \\}

The reinforcement learning (RL) agent can built using various RL algorithms such as Soft Actor-Critic (SAC), Proximal Policy Optimization (PPO), \dots, which are implemented in the \texttt{stable-baselines3} library \cite{stable-baselines3-sac}. For this specific task, we employed the \textbf{Soft Actor-Critic (SAC)} algorithm~\cite{haarnoja2018soft}, which is well-suited for handling the continuous and high-dimensional action space of the beamline environment. SAC is known for its sample efficiency, stability, and ability to explore large action spaces. The use of off-policy learning and maximum entropy principles further improves the exploration and robustness of the learned policy.

Initially, SAC was implemented using the \texttt{stable-baselines3} library~\cite{stable-baselines3-sac}, which provided compatibility with our custom-designed environment and allowed for easy experimentation with various RL algorithms. Later, to gain more control over the agent's architecture, hyperparameters, and training process, we re-implemented the SAC algorithm from scratch. This custom implementation enabled fine-tuning of the agent’s behavior and allowed for a deeper understanding of its interaction with the beamline environment.

By combining the flexibility of \texttt{stable-baselines3} with our custom SAC implementation, we established a versatile framework that facilitates experimentation with different optimization strategies, offering both flexibility and precision tailored to the specific needs of accelerator control.

\section{\uppercase{Numerical Experiments}}
\label{sec:Numerical Experiments}

To evaluate the performance of our reinforcement learning framework, we conducted two experiments using the same beamline structure which consists of seven quadrupole magnets, and the optimization objective in both cases was to deliver the maximum number of particles to the target. The experiments differ in their initial beam emittance, which is a fundamental property of particle beams. Emittance reflects the phase space area occupied by the particles, incorporating both the spatial distribution (beam size) and angular spread (momentum distribution) of the particles. A well-collimated beam has low emittance, and in an ideal scenario without losses, the emittance remains conserved throughout the beamline.

\bigskip

\subsection{Experiment 1: Nominal Beam Emittance\\ \\}

\begin{figure}[p]
\centerline{\includegraphics[width=0.75\linewidth]{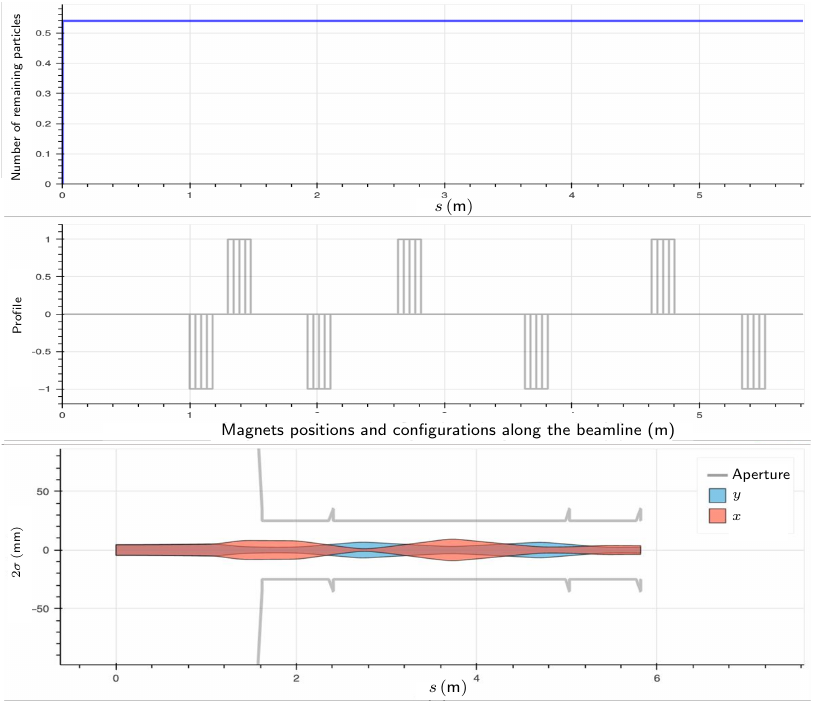}}
\vspace{-0.5\baselineskip}
        \caption{Beam profile at the target for Experiment 1. Emittance = 2000. The top panel shows a stable particle count along the beamline, indicating 100\% particle delivery. The middle panel shows the magnet current profiles, while the bottom panel illustrates the particle distribution along \( x \) and \( y \) axes within the aperture.}
        \label{fig:result_2000}

\medskip

\centerline{\includegraphics[width=0.75\linewidth]{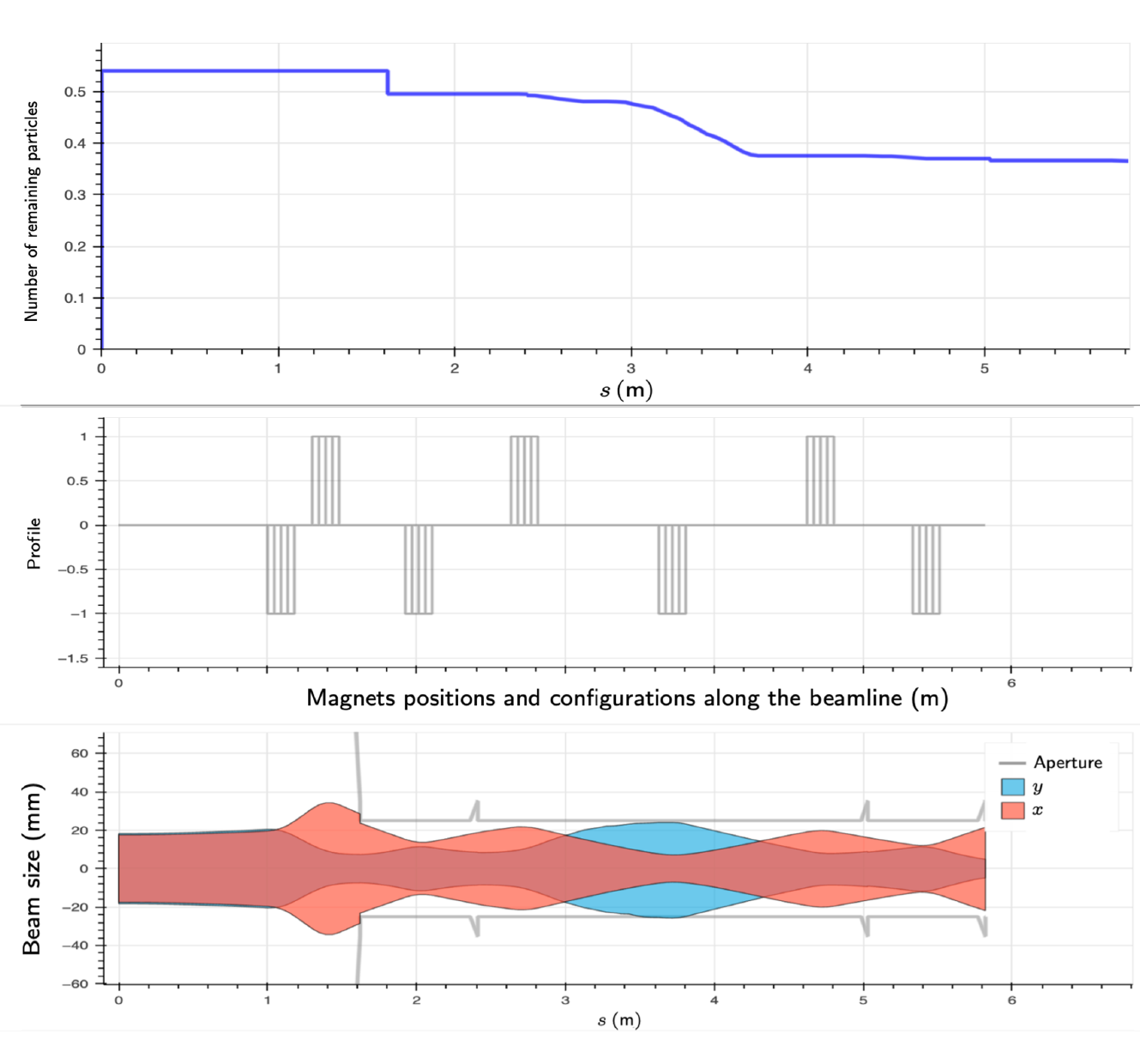}}
\vspace{-0.5\baselineskip}
        \caption{Beam profile at the target for Experiment 2. Emittance = 32,000. The top panel depicts the particle count along the beamline, indicating a 32.4\% particle loss. The middle panel shows the optimized magnet current profiles, and the bottom panel shows the particle distribution on the \( x \) and \( y \) axes, with the beam staying mostly within the aperture.}
        \label{fig:result_32000}
\end{figure}

In the first experiment, the beam had a nominal emittance of 2000. The reinforcement learning agent, employing the Soft Actor-Critic (SAC) algorithm, optimized the currents of the quadrupole magnets to maximize particle delivery. As shown in Figure ~\ref{fig:result_2000}, the agent successfully delivered all particles to the target, achieving 100\% transmission efficiency. 

This experiment highlights the RL agent's ability to achieve optimal performance under ideal beam conditions.

\bigskip

\subsection{Experiment 2: Increased Beam Emittance\\ \\}

The second experiment involved the same beamline configuration but with an increased beam emittance of 32,000, sixteen times higher than in the first experiment. A larger emittance represents a ``messier'' beam, with particles exhibiting a broader spatial and angular distribution, significantly complicating the optimization task.


Despite the challenges posed by this degraded beam quality, the RL agent adapted effectively to the new conditions. By optimizing the quadrupole magnet currents, the agent successfully delivered 67.6\% of the particles to the target, losing 32.4\% of the beam in the process. Figure~\ref{fig:result_32000} visualizes these results. 

This experiment underscores the robustness of the RL framework in managing more complex beam conditions. The ability to mitigate particle losses and deliver the majority of the beam to the target demonstrates the versatility and effectiveness of the SAC algorithm in optimizing accelerator operations under adverse scenarios.

\label{sec:Conclusion} 
\section{\uppercase{Conclusion}}

In this study, we demonstrate the efficacy of reinforcement learning (RL) in optimizing accelerator control through a simulation-based approach. By integrating the Python-based RL framework with high-fidelity simulation tools like \texttt{Elegant}, we develop a robust environment for training RL agents tailored to beamline optimization. Our experiments illustrate the flexibility of the proposed framework in addressing diverse beam configurations, achieving 100\% transmission efficiency for nominal emittance and 67.6\% for significantly higher emittance, thereby highlighting the adaptability of the RL agent to challenging conditions. This work showcases the potential of RL to enhance accelerator performance by reducing manual tuning efforts, improving transmission efficiency, and supporting complex optimization tasks. Future directions include applying this framework to multi-objective optimization, incorporating real-time feedback from experimental facilities, and extending its applicability to other accelerator designs and control problems.

\newpage

\section*{\uppercase{Acknowledgments}}

This work is supported by HSE Basic research fund. The computation for this research was performed using the computational resources of HPC facilities at HSE University.

\section*{\uppercase{Conflict of interest}}
The authors declare that they have no conflicts of interest.

\bibliographystyle{pepan}
\bibliography{I02}

\end{document}